\documentclass{ws-procs9x6}

\def\be{\begin{eqnarray}}
\def\ee{\end{eqnarray}}
\def\beq{\begin{equation}}
\def\eeq{\end{equation}}
\def\ba{\begin{array}}
\def\ea{\end{array}}
\def\hc{{\rm H.c.}}
\def\bec{\begin{center}}
\def\ec{\end{center}}

\begin{document}

\title{PARTIALLY COMPOSITE TWO HIGGS DOUBLET MODEL}

\author{P. KO}

\address{School of Physics, KIAS \\
Seoul 130-722, Korea \\
$^*$E-mail: pko@kias.re.kr} 

\begin{abstract}
We consider a possibility that electroweak symmetry breaking (EWSB) is 
triggered by a fundamental Higgs and a composite Higgs arising 
in a dynamical symmetry breaking mechanism induced by a new strong dynamics. 
The resulting Higgs sector is a partially composite two-Higgs doublet model 
with specific boundary conditions on the coupling and mass parameters 
originating at a compositeness scale $\Lambda$.  
The phenomenology of this model is discussed including the collider 
phenomenology at LHC and ILC.
\end{abstract}


\bodymatter

\section{Introduction}\label{sec1:intro}

EWSB is the origin of  the masses of chiral fermions and electroweak 
gauge bosons as well as CP violation in the quark sector within the 
standard model (SM).
It is most important in particle physics to understand the origin of EWSB, 
and LHC will serve for this purpose. 
There have been many attempts to construct interesting models 
for EWSB beyond the SM \cite{hillsimmons}. 

Dynamical symmetry breaking \'{a} la Miransky, Tanabashi, Yamawaki (MTY)   
\cite{MTY} and Bardeen, Hill, Lindner (BHL)
\cite{BHL} is a particularly interesting scenario, 
since the heavy top mass is intimately related with a new strong dynamics 
that condenses the $t\bar{t}$ bilinear, 
and breaks the EW symmetry down to $U(1)_{\rm EM}$.
Both heavy top mass and Higgs mass are generated dynamically, in anology 
with superconductivity of Bardeen-Cooper-Schrieffer (BCS). 

However, there are basically two drawbacks in this model. First, 
the origin of the new strong interactions that triggers EWSB is not clear. 
The attractive 4-fermion interaction is simply put in by hand 
within the BHL model. 
Also, the original version of BHL with 3 families or its extension with 
two Higgs doublets \cite{luty} predict that the top mass should be 
significantly heavier than the experimental observation: 
$m_t = 170.9 \pm 1.1 ({\rm stat}) \pm 1.5 ({\rm sys)}$ GeV \cite{topmass}.
However, these two drawbacks could be evaded within extra 
dimensional scenarios, without ruining its niceties. 

If QCD is a bulk theory, then it is possible that the KK gluon exchange 
can induce attractive Nambu-Jona-Lasinio (NJL) type four-fermion 
interaction in the low energy regime, and dynamical 
symmetry breaking can occur in a natural way \cite{dobrescu1}. 
It should be emphasized that this is completely different from another 
popular way of symmetry breaking in extra dimension, namely symmetry 
breaking by boundary conditions. 
Therefore, in extra dimensional scenarios, electroweak symmetry 
can be broken by fundamental Higgs, by boundary condition or 
by some dynamical mechanism. 
Generically all three possibilities could be present altogether. 
In most recent studies, the gauge symmetries were broken by the nontrivial 
boundary conditions with or without fundamental Higgs.
In this talk, I discuss another possibility, where electroweak symmetry 
is broken by fundamental Higgs VEV's, as well as dynamically by $t\bar{t}$ 
condensate \cite{pko}.  This way we will find that we can avoid both 
drawbacks of BHL model.


\section{A Model of Dynamical EWSB  with a Fundamental Scalar}

Our model is a simple extension of the SM. We assume there is  a new 
strong dynamics at some high energy scale  $\Lambda$, which is  
effectively described by the NJL type four-fermion interaction term:
\be
{\cal L} = {\cal L}_{\rm SM} +
      G ( \overline{\psi}_{L} t_{R} )( \overline{t}_{R} \psi_{L} ) ,
\ee
where the SM Higgs doublet $\phi$ is included from the beginnig, 
unlike the BHL model.
The explicit forms of the SM lagrangians can 
be found in Ref.~\cite{pko}.
The Yukawa couplings for the 1st and the 2nd generations 
do not play any roles in our analysis, and will be ignored  in the following.
We don't specify the origin of this NJL type interaction, but 
the KK gauge boson exchange in extra dimension scenarios could be one 
possible origin of this new strong interaction.
As a minimal extension of the SM, we assume that this new strong dynamics 
acts only (or dominantly) on top quark. 

We can rewrite the NJL term in Eq.(2.1) in terms of an auxiliary scalar 
field $\Phi$: 
\be
{\cal L} = {\cal L}_{\rm SM}
           + g_{t0} ( \overline{\psi}_{L} t_{R} \tilde \Phi + \hc )
           -M^2 \Phi^{\dagger} \Phi,
\ee
where $G = g_{t0}^2 / M^2$ with $M \sim \Lambda$.
$g_{t0}$ is a newly defined Yukawa coupling
between the top quark and the auxiliary scalar field $\Phi$.
$\Phi$ describes the composite scalar bosons that appear 
when the $\langle \bar{t} t \rangle$ develops nonvanishing VEV and breaks 
the electroweak symmetry. 
Far below the scale $\Lambda$, the $\Phi$ field will develop the kinetic 
term due to quantum corrections  and become dynamical. 
The resulting low energy effective field theory will
be two-Higgs doublet model, one being a fundamental Higgs $\phi$ and the 
other being a composite Higgs $\Phi$.  
Thus it can be called a partially composite two-Higgs  doublet (PC2HD) model.  

In order to avoid too large FCNC mediated by neutral Higgs bosons,
we assign a $Z_2$ discrete symmetry under which 
the lagrangian is invariant ;
\begin{equation} 
( \Phi,~\psi_L,~U_R )  \to  + ( \Phi,~\psi_L,~U_R ) , ~~~
( \phi , D_R )  \to  - ( \phi, D_R ) .
\end{equation}
With this $Z_2$ discrete symmetry, $t$ and $b$ couple to $\Phi$ and the
SM Higgs, respectively.
In consequence, our model becomes the Type-II two-Higgs doublet model 
as the minimal supersymmetric standard model (MSSM).

The renormalized lagrangian for the scalar fields at low energy 
is given by
\be
{\cal L}_{\rm ren} &=& Z_\phi(D_\mu\phi)^\dagger(D^\mu \phi)
          + Z_\Phi (D_\mu \Phi)^\dagger(D^\mu \Phi)
          - V(\sqrt{Z_\phi}\phi, \sqrt{Z_\Phi}\Phi)
\nonumber \\
  &&~~~+ \sqrt{Z_\Phi}g_t(\overline{\psi}_L t_R \tilde \Phi + {\rm h.c})
          + \sqrt{Z_\phi}g_b(\overline{\psi}_L b_R \phi + {\rm h.c}),
\ee
with
\be
V(\phi, \Phi) &=& \mu_1^2\phi^\dagger\phi + \mu_2^2\Phi^\dagger \Phi
              + ( \mu_{12}^2  \phi^\dagger \Phi + \hc)
              + \frac{1}{2}\lambda_1(\phi^\dagger\phi)^2
+ \frac{1}{2}\lambda_2(\Phi^\dagger \Phi)^2
\nonumber \\
         &&~+ \lambda_3(\phi^\dagger\phi)(\Phi^\dagger \Phi)
              + \lambda_4|\phi^\dagger \Phi|^2
              + \frac{1}{2} [ \lambda_5 (\phi^\dagger \Phi)^2 + \hc ],
\ee
In the scalar potential, we have introduced a dimension-two  
$\mu_{12}^2$ term that breaks the discrete symmetry softly  
in order to generate the nonzero mass for the CP-odd Higgs boson. 
Otherwise the CP-odd Higgs boson $A$ would be an unwanted axion related
with spontanesously broken global $U(1)$ Peccei-Quinn symmetry, 
which  would be a phenomenological disaster.
This $\mu_{12}^2$ parameter will be traded with the 
$m_A^2$, the (mass)$^2$ parameter of the CP-odd Higgs boson, 
which is another free parameter of our model.

Matching the lagrangian with Eq. (2.5) at the compositeness scale $\Lambda$,
we obtain the following matching conditions as $\mu \rightarrow \Lambda$:
\be
&& \sqrt{Z_\phi}\rightarrow 1,~~~~~ \sqrt{Z_\Phi} \rightarrow 0,
~~~~~Z_\phi \mu_1^2 \rightarrow m_{0}^2,~~~~~
Z_\Phi \mu_2^2 \rightarrow M^2,
\nonumber  \\
&& Z_\phi \lambda_1 \rightarrow \lambda_{10},~~~~~
Z_\Phi^2\lambda_2 \rightarrow 0,~~~~~
Z_\phi Z_\Phi \lambda_{i=3,4,5} \rightarrow 0.
\ee
These conditions are the boundary conditions for the RG equations. 

Before proceeding, we would like to compare our model with Luty's model
\cite{luty}.  In Luty's model, both Higgs doublets are composite, and thus 
the matching conditions for $Z_\phi$ and $\lambda_1$ become 
\begin{equation}
\sqrt{Z_\phi}\rightarrow 0,~~~~~ 
Z_\phi \lambda_1 \rightarrow 0,~~~~~
\end{equation}
which are different from those in our model.
These different matching conditions lead to very different predictions 
for the scalar boson spectra compared to the Luty's model. 
Also we have additional Yukawa coupling $g_b$ so that we can fit both 
the bottom and the top quark masses without difficulty unlike the models
by BHL or Luty.

\section{Particle Spectra and predictions}

Our model is defined in terms of three parameters: Higgs self coupling 
$\lambda_{10}$, the compositeness scale $\Lambda$ where $\lambda_{10}$ and 
the NJL interaction are specified, and the CP-odd Higgs boson mass $m_A$. 
Since $\lambda_{10}$ is also present in the SM, our model has two more 
parameters compared with the SM. 
It is strightforward to analyze the conditions for the correct EWSB and 
the particle spectra. The details can be found in the original paper
\cite{pko}. In the following, I highlight the main results of our model:
\begin{itemize}
\item We can fit both the top and the bottom masses without difficulty in 
our model, unlike the BHL model or the Luty model, since the bottom quark
get massive due to the fundamental Higgs. The allowed region of 
$\tan\beta$ is rather narrow: $0.45 \lesssim \tan\beta \lesssim 1$. 
Therefore the $W$ and the $Z$ boson get their masses almost equally from 
the fundamental Higgs and the $t\bar{t}$ condensation in our model.
\item Since $\tan\beta \lesssim 1$, there is a strong constraint from 
$B\rightarrow X_s \gamma$, which implies that the charged Higgs boson 
should be heavier than $\sim 400$ GeV. 
\item There is no CP violating mixing in the neutral Higgs boson sector, 
since $\lambda_5$ remains zero at all scale within our model.
\item $m_H^\pm \lesssim m_A$ in our model, and the charged Higgs can be 
even lighter than the lightest neutral Higgs boson, when the composite 
scale $\Lambda$ is high. See Fig.~1.
\item Triple and quartic self couplings of Higgs bosons can deviate 
from the SM values by significant amounts.
\item Higgs coupling to the top quark is enhanced in our model so that
the Higgs production rate at LHC is larger than the SM values. 
On the other hand, the Higgs productions at the ILC through 
Higgs--stralhlung and the $WW$ fusion are suppressed compared to 
the SM values. See Fig.~2.
\end{itemize}

\section{Conclusions}

In this talk, I considered a possibility that the Higgs 
boson produced at the future colliders is neither a fundamental scalar
nor a composite scalar, but a mixed state of them.
It could be a  generic feature, if there exists a strong dynamics
at a high scale which give rise to the dynamical electroweak symmetry 
breaking, in addition to the usual Higgs mechanism due to the 
nonvanishing  VEV  of a fundamental Higgs. 
It is interesting that this scenario could be easily realized,  
if we embed the SM lagrangian in a higher dimension with bulk gauge 
interactions.
The resulting theory can accommodate the observed top mass, and give 
specific predictions for neutral and charged Higgs masses at a given value
of $\Lambda$. 
Whether such scenario is realized or not in nature could be studied at
LHC and ILC.


\section*{Acknowledgments}

The author is grateful to B. Chung, D.W. Jung and K.Y. Lee for enjoyable
collaborations, and Prof. Yamawaki for discussions and his wonderful 
organization of the workshop.  
This work is supported in part by KOSEF through CHEP at Kyungpook National 
University.

\def\PRD #1 #2 #3 {{\em Phys. Rev. D} {\bf#1},\ #2 (#3)}
\def\PRL #1 #2 #3 {{\em Phys. Rev. Lett.} {\bf#1},\ #2 (#3)}
\def\PLB #1 #2 #3 {{\em Phys. Lett. B} {\bf#1},\ #2 (#3)}
\def\NPB #1 #2 #3 {{\em Nucl. Phys.} {\bf B#1},\ #2 (#3)}
\def\ZPC #1 #2 #3 {{\em Z. Phys. C} {\bf#1},\ #2 (#3)}
\def\EPJ #1 #2 #3 {{\em Euro. Phys. J. C } {\bf#1},\ #2 (#3)}
\def\JHEP #1 #2 #3 {{\em JHEP} {\bf#1},\ #2 (#3)}
\def\IJMP #1 #2 #3 {{\em Int. J. Mod. Phys. A} {\bf#1},\ #2 (#3)}
\def\MPL #1 #2 #3 {{\em Mod. Phys. Lett. A} {\bf#1},\ #2 (#3)}
\def\PTP #1 #2 #3 {{\em Prog. Theor. Phys.} {\bf#1},\ #2 (#3)}
\def\PR #1 #2 #3 {{\em Phys. Rep.} {\bf#1},\ #2 (#3)}
\def\RMP #1 #2 #3 {{\em Rev. Mod. Phys.} {\bf#1},\ #2 (#3)}
\def\PRold #1 #2 #3 {{\em Phys. Rev.} {\bf#1},\ #2 (#3)}
\def\IBID #1 #2 #3 {{\it ibid.} {\bf#1},\ #2 (#3)}


\begin{figure}[ht]
\begin{center}
\hbox to\textwidth{\hss\epsfig{file=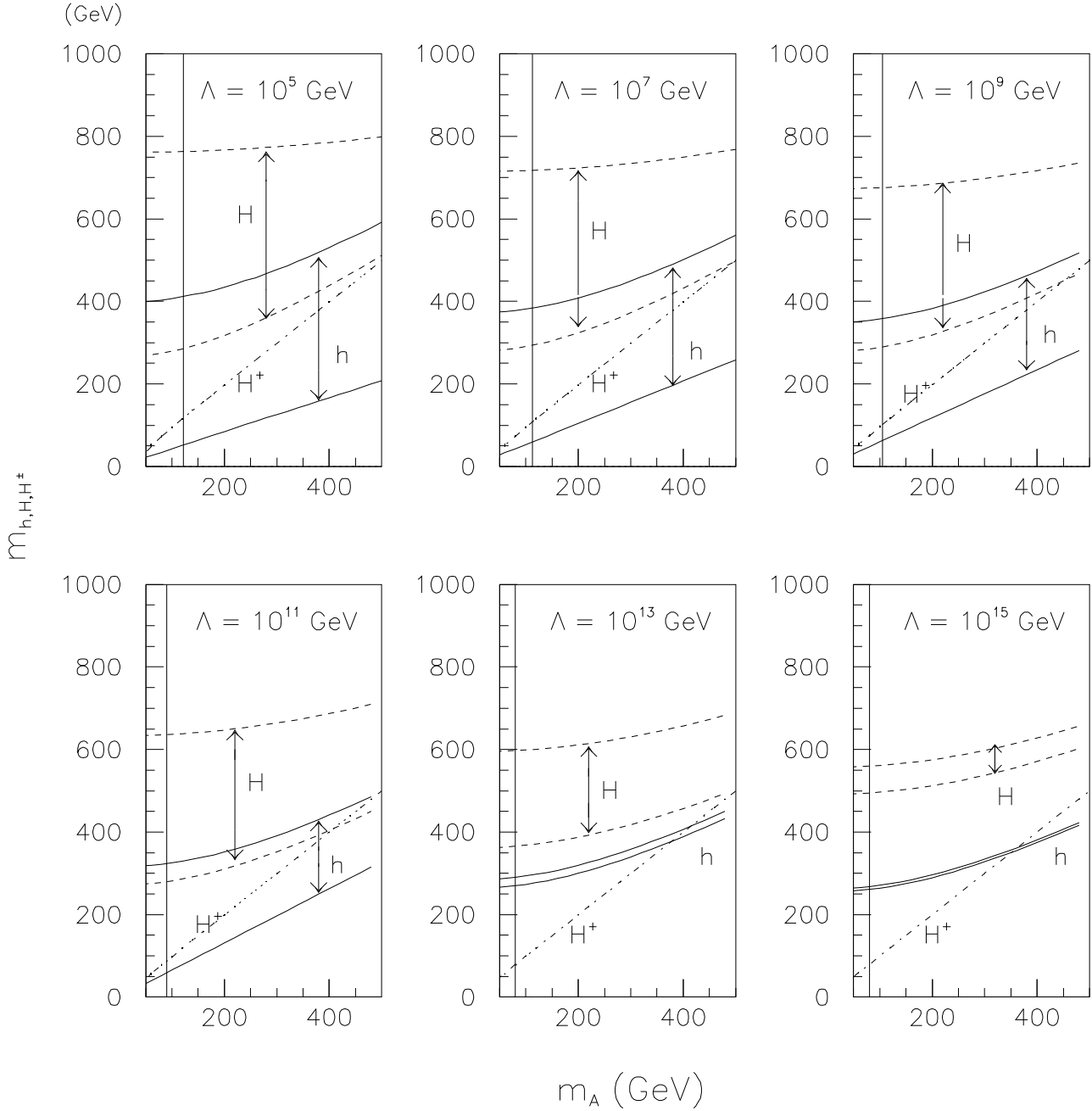,height=18.5cm}\hss}
\vspace{0.2cm}
\caption{
Masses of neutral Higgs bosons $h$ (inside the solid lines), 
$H$ (inside the dashed lines) and the charged Higgs boson $H^\pm$ 
(dahs-dotted line) with respect to $m_A$.
}
\end{center}
\end{figure}

\begin{figure}[ht]
\begin{center}
\hbox to\textwidth{\hss\epsfig{file=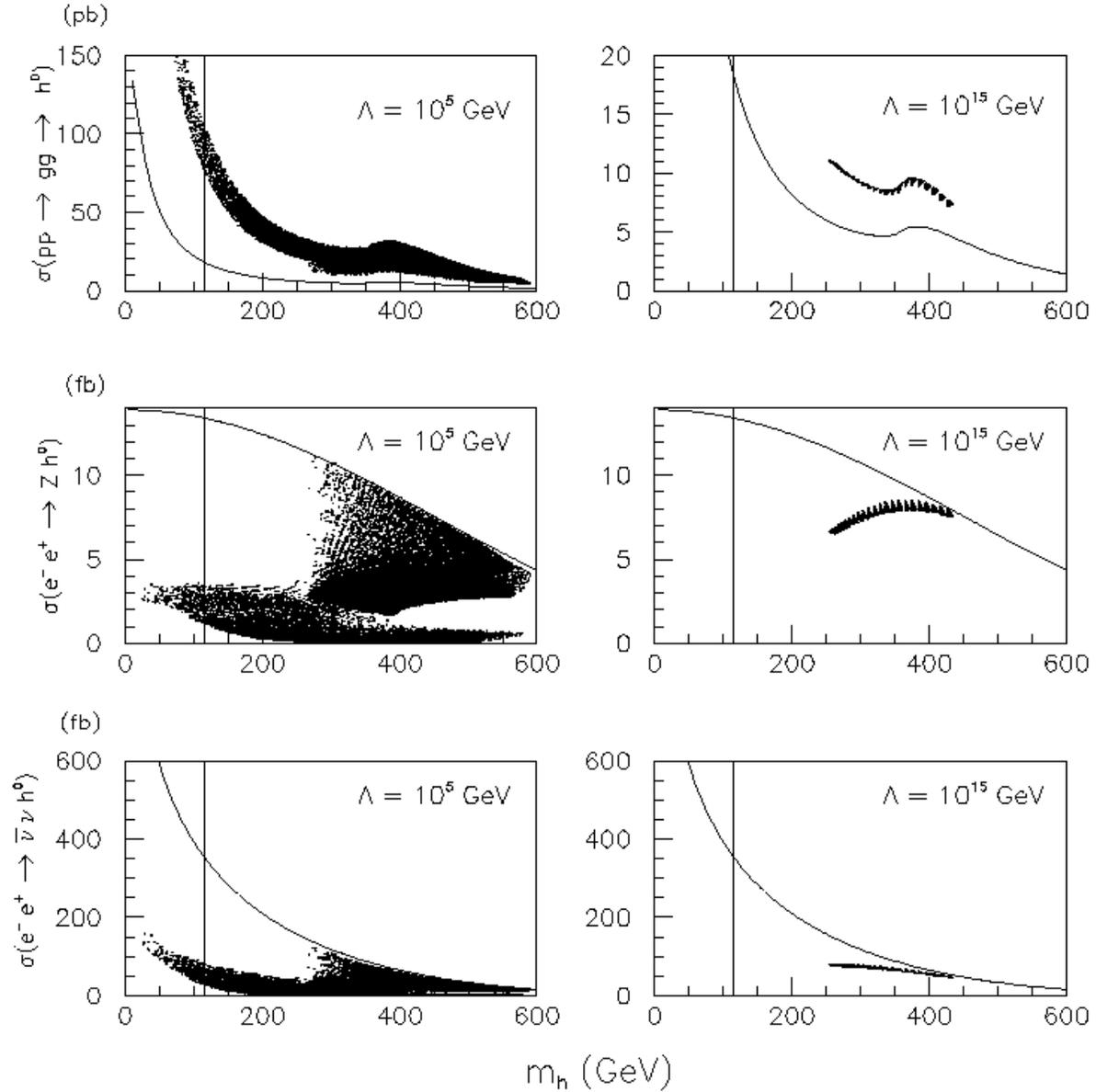,width=17cm}\hss}
\vspace{0.2cm}
\caption{
Production  cross section of the neutral Higgs boson at 
the LHC and ILC. $\sqrt{s} = 14$ TeV for the LHC and  $\sqrt{s} = 1$ TeV for
the ILC are assumed.
The solid curves denotes the SM predictions.
}
\end{center} 
\end{figure}

\end{document}